\documentclass[3p,times,procedia]{elsarticle}
\usepackage{nupha_ecrc}

\volume{00}
\firstpage{1}
\journalname{Nuclear Physics A}
\runauth{R.-A. Tripolt et al.}
\jid{nupha}
\jnltitlelogo{Nuclear Physics A}

\usepackage{amssymb}
\usepackage[figuresright]{rotating}
\usepackage{slashed}

\usepackage{etex}
\usepackage{graphicx}
\usepackage{xspace}
\usepackage{xcolor}
\usepackage[absolute,overlay]{textpos}
\usepackage{ulem}
\usepackage{tikz}
\usepackage{amsmath}
\usepackage{dsfont}
\usepackage{framed}
\usepackage{wasysym}
\usepackage[scaled=1.03]{inconsolata}
\usepackage{tabularx}
\usepackage{geometry}
\usepackage{transparent}

\graphicspath{
	{./}
	{./figures/}
}

\newcommand{\Tr}{\rm{Tr}}

\begin{document}

\begin{frontmatter}

\dochead{XXVIIth International Conference on Ultrarelativistic Nucleus-Nucleus Collisions\\ (Quark Matter 2018)}

\title{In-medium spectral functions and dilepton rates with the Functional Renormalization Group }

\author[ECT]{Ralf-Arno Tripolt}
\author[JLU]{Christopher Jung}
\author[ECT]{Naoto Tanji}
\author[JLU]{Lorenz von Smekal}
\author[ECT,TUDa]{Jochen~Wambach}
\address[ECT]{European Centre for Theoretical Studies in Nuclear Physics and related Areas (ECT*) and Fondazione Bruno Kessler, Villa Tambosi, 38123 Villazzano (TN), Italy}
\address[JLU]{Institut f\"ur Theoretische Physik, Justus-Liebig-Universit\"at, 35392 Giessen, Germany}
\address[TUDa]{Theoriezentrum, Institut f\"ur Kernphysik, Technische Universit\"at Darmstadt, 64289 Darmstadt, Germany}

\begin{abstract}
We present recent results on in-medium spectral functions of vector and axial-vector mesons, the electromagnetic (EM) spectral function and dilepton rates using the Functional Renormalization Group (FRG) approach. Our method is based on an analytic continuation procedure that allows us to calculate real-time quantities like spectral functions at finite temperature and chemical potential. As an effective low-energy model for Quantum Chromodynamics (QCD) we use an extended linear-sigma model including quarks where (axial-)vector mesons as well as the photon are introduced as gauge bosons. In particular, it is shown how the $\rho$ and the $a_1$ spectral function become degenerate at high temperatures or chemical potentials due to the restoration of chiral symmetry. Preliminary results for the EM spectral function and the dilepton production rate are discussed with a focus on the possibility to identify signatures of the chiral crossover and the critical endpoint (CEP) in the QCD phase diagram.
\end{abstract}


\end{frontmatter}

\section{Introduction}
\label{intro}

Exploring the phase structure of QCD, locating the critical endpoint and describing the in-medium modifications of hadrons are central challenges in the focus of ongoing experimental and theoretical efforts. The necessary energy densities can be reached in heavy-ion collisions, where electromagnetic probes such as photons and dileptons, as decay products of vector mesons, are uniquely well-suited to investigate these questions  \cite{RappWambachHees2010}. 

In this work we present first results for the EM spectral function and the dilepton rate based on the theoretical framework for the calculation of spectral functions proposed in \cite{Kamikado2014, Tripolt2014,Tripolt2014a,TripoltSmekalWambach2016a} which was applied to  vector and axial-vector mesons in \cite{JungRenneckeTripoltEtAl2017}. Within this framework the FRG is used in the following, which represents a non-perturbative continuum method that is well-suited to describe phase transitions. It is applied to an extended linear-sigma model including quarks, which serves as a low-energy model of QCD, albeit without incorporating confinement.

\section{Flow equations for vector mesons}
\label{flow}

\begin{figure}[h]
	\includegraphics[width=\textwidth]{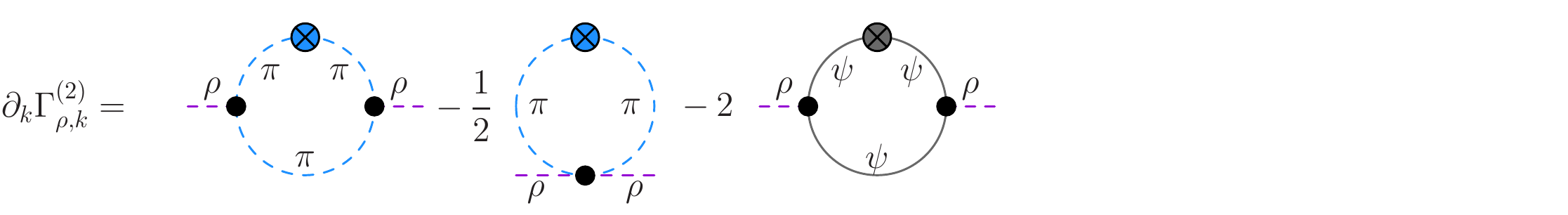}\\
	\includegraphics[width=\textwidth]{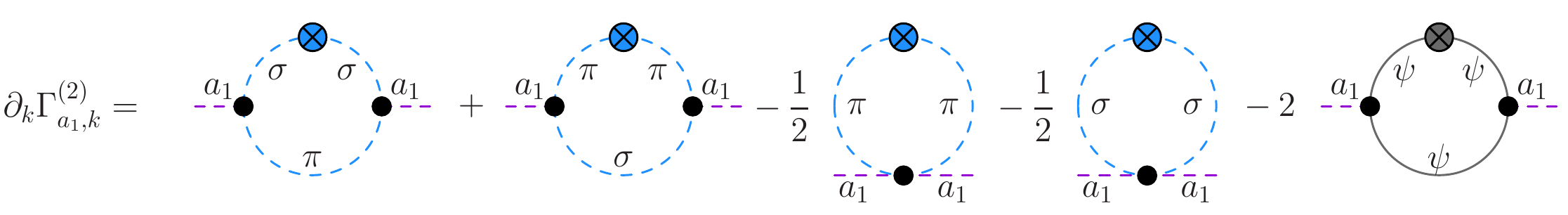} 
	\caption{Diagrammatic representation of the flow equations for the $\rho$ and the $a_1$ 2-point function. The vertices, which are taken to be momentum-independent, are represented by black dots while the regulator insertions, $\partial_k R_k$, are represented by crossed circles.}
	\label{fig:flow}
\end{figure}

We use the same theoretical framework as discussed in \cite{JungRenneckeTripoltEtAl2017}. As an ansatz for the effective average action we choose
\begin{eqnarray}
\Gamma_k &=& \int d^4x \bigg\{\bar{\psi} \left(\slashed{D}-\mu\gamma_0+
h_{S}\left(\sigma +\mathrm{i} \vec{\tau}\vec{\pi}\gamma_5\right)+
\mathrm{i} h_{V} \left(\gamma_{\mu} \vec{\tau}\vec{\rho}^{\mu}+\gamma_{\mu}\gamma_5 \vec{\tau}\vec{a}_1^{\mu}\right)
\right)\psi+ U_{k}(\phi^2)
-c\sigma\nonumber \\
&&\hspace{4cm} +\frac{1}{2} \left|(D_{\mu}-\mathrm{i}gV_\mu)\phi\right|^2 
+\frac{1}{8} \Tr\left(V_{\mu\nu} V^{\mu\nu}\right)
+\frac{1}{4}m_{V,k}^2 \Tr \left(V_{\mu}V^{\mu}\right)
\bigg\}
\end{eqnarray}
with $V_{\mu\nu} = D_\mu V_\nu-D_\nu V_\mu-\mathrm{i}g\left[V_\mu,V_\nu\right]$, the covariant derivatives
$D_\mu \psi = \left(\partial_\mu-\mathrm{i}eA_\mu Q\right)\psi$ and
$D_\mu V_\mu = \partial_\mu V_\nu-i e A_\mu [T_3,V_\nu]$, the scalar sigma and pseudo-scalar pion fields
$\phi\equiv (\vec{\pi},\sigma)$ and the vector and axial-vector meson fields
$V_\mu\equiv \vec{\rho}_\mu \vec{T}+\vec{a}_{1,\mu} \vec{T}^5$. This represents a low-energy model of two-flavor QCD which incorporates the effects of chiral symmetry. The $\rho$ and the $a_1$ field as well as the photon field $A_\mu$ are introduced using the gauging principle, see also \cite{1960AnPhy,Lee:1967ug}. The flow equations for the $\rho$ and the $a_1$ 2-point functions are analytically continued to real energies $\omega$, using the method described in \cite{JungRenneckeTripoltEtAl2017}, and then solved numerically. In this work we will use the Vector Meson Dominance (VMD) assumption and neglect decay channels into vector mesons, cf.~\cite{JungRenneckeTripoltEtAl2017}, which yields the diagrammatic structure of the flow equations shown in Fig.~\ref{fig:flow}. For the implementation of fluctuating vector mesons within the FRG we refer to \cite{JungSmekal}.

\section{EM Spectral function and dilepton rates}
\label{dilepton}

\begin{figure}[t]
	\hspace{3cm}\includegraphics[width=\textwidth]{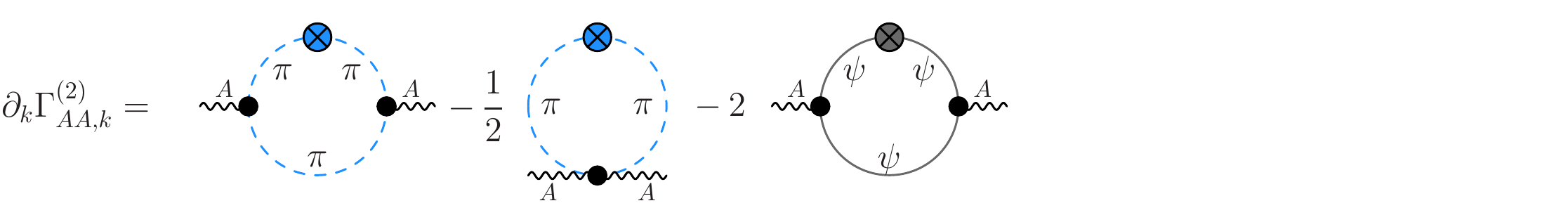}\\
	\vspace{0mm}
	\hspace{3cm}\includegraphics[width=\textwidth]{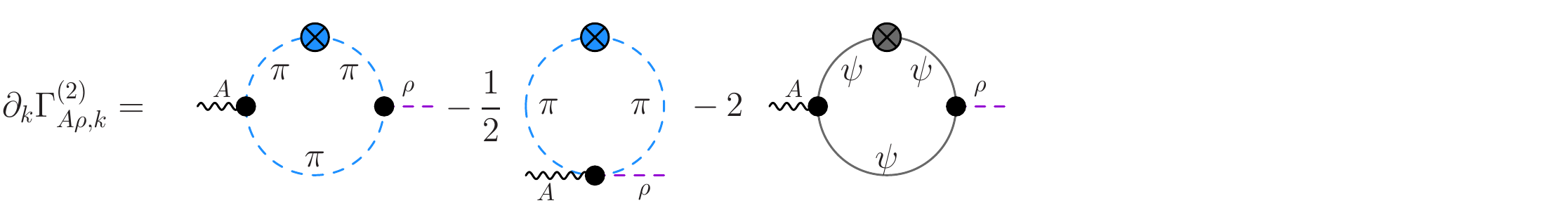}
	\caption{Diagrammatic representation of the flow equations for the bare photon and the mixed rho-photon 2-point function in the VMD approximation, \cite{JungTanjiTripoltEtAl}. After diagonalization, the full photon 2-point function can be extracted, see also Eq.~(\ref{eq:matrix}).}
	\label{fig:EMflow}
\end{figure}

Since the $\rho$ vector meson mixes with the photon field $A_\mu$, one has to diagonalize the following matrix of 2-point functions in order to extract the full EM spectral function connected to $\tilde{\Gamma}_{AA}^{(2)}$:
\begin{equation}
\label{eq:matrix}
\hspace{-5mm}\begin{pmatrix}
\Gamma^{(2)}_{AA} & \Gamma^{(2)}_{A\rho}\\
\Gamma^{(2)}_{\rho A} & \Gamma^{(2)}_{\rho\rho}
\end{pmatrix}
\hspace{1mm}\underrightarrow{\rm{diagonalize}}\hspace{1mm}
\begin{pmatrix}
\tilde{\Gamma}^{(2)}_{AA} & 0\\
0 & \tilde{\Gamma}^{(2)}_{\rho\rho}
\end{pmatrix},\quad
\tilde{\Gamma}^{(2)}_{AA}=
\Gamma^{(2)}_{AA}-\frac{\Gamma^{(2)}_{A\rho}\Gamma^{(2)}_{\rho A}}{\Gamma^{(2)}_{\rho\rho}}+ \mathcal{O}(e^4).
\end{equation}
For details on the derivation of these flow equations, which are shown diagrammatically in Fig.~\ref{fig:EMflow}, we refer to \cite{JungTanjiTripoltEtAl}. 

\begin{figure}[b!]
	{\includegraphics[width=0.45\textwidth]{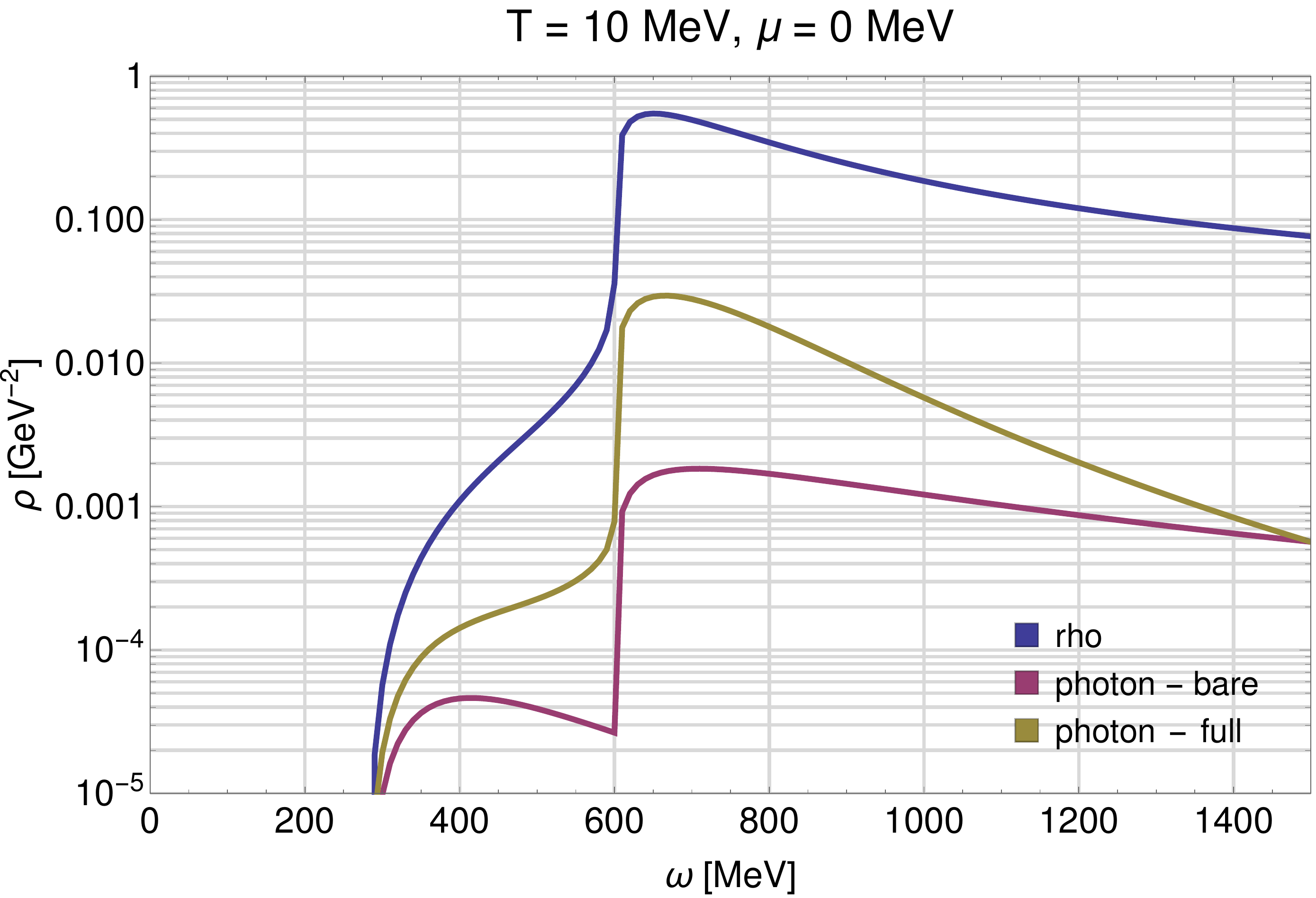}}
	\hspace{6mm}
	{\includegraphics[width=0.45\textwidth]{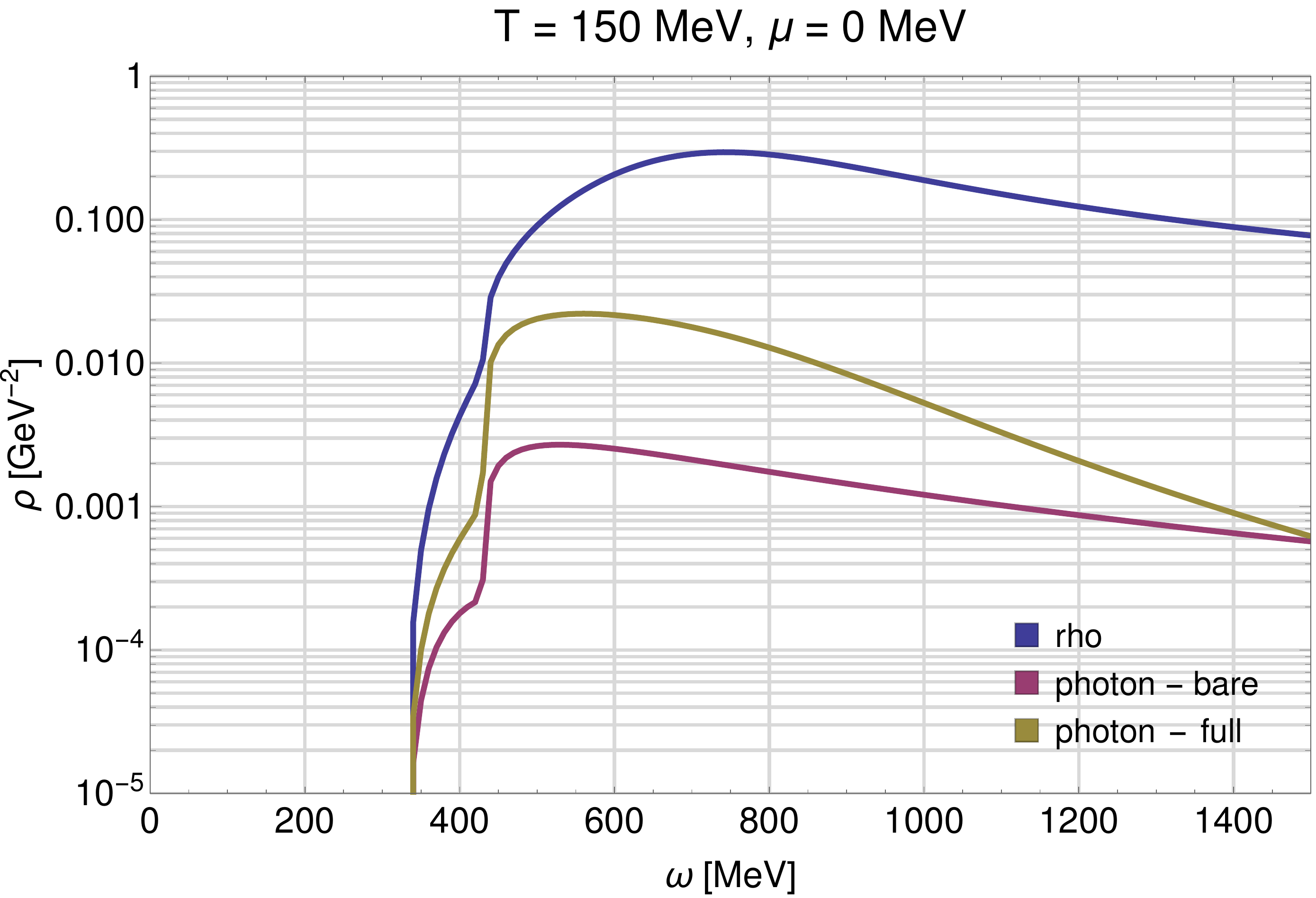}}\\
	\vspace{3mm}
	{\includegraphics[width=0.45\textwidth]{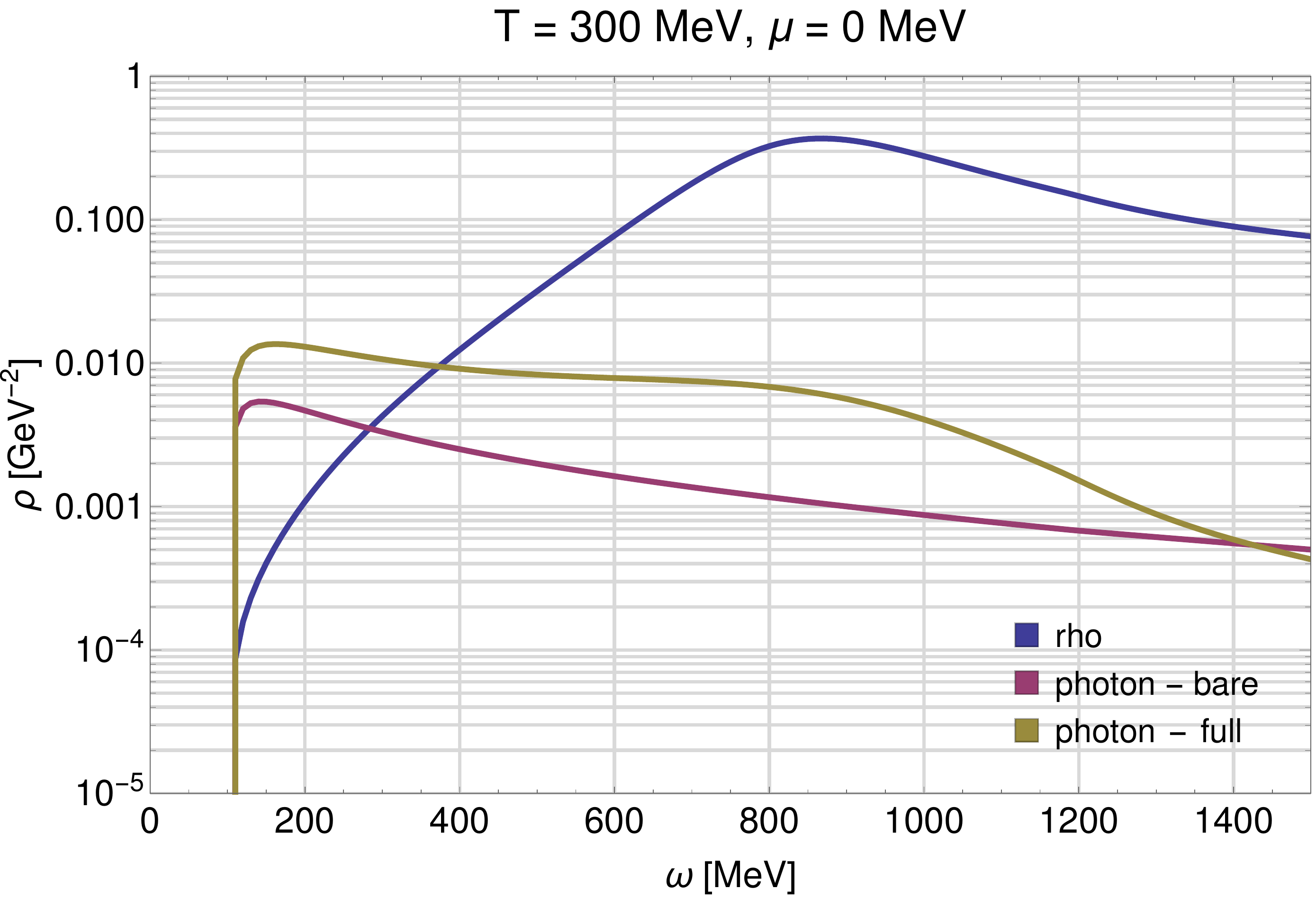}}
	\hspace{6mm}
	{\includegraphics[width=0.45\textwidth]{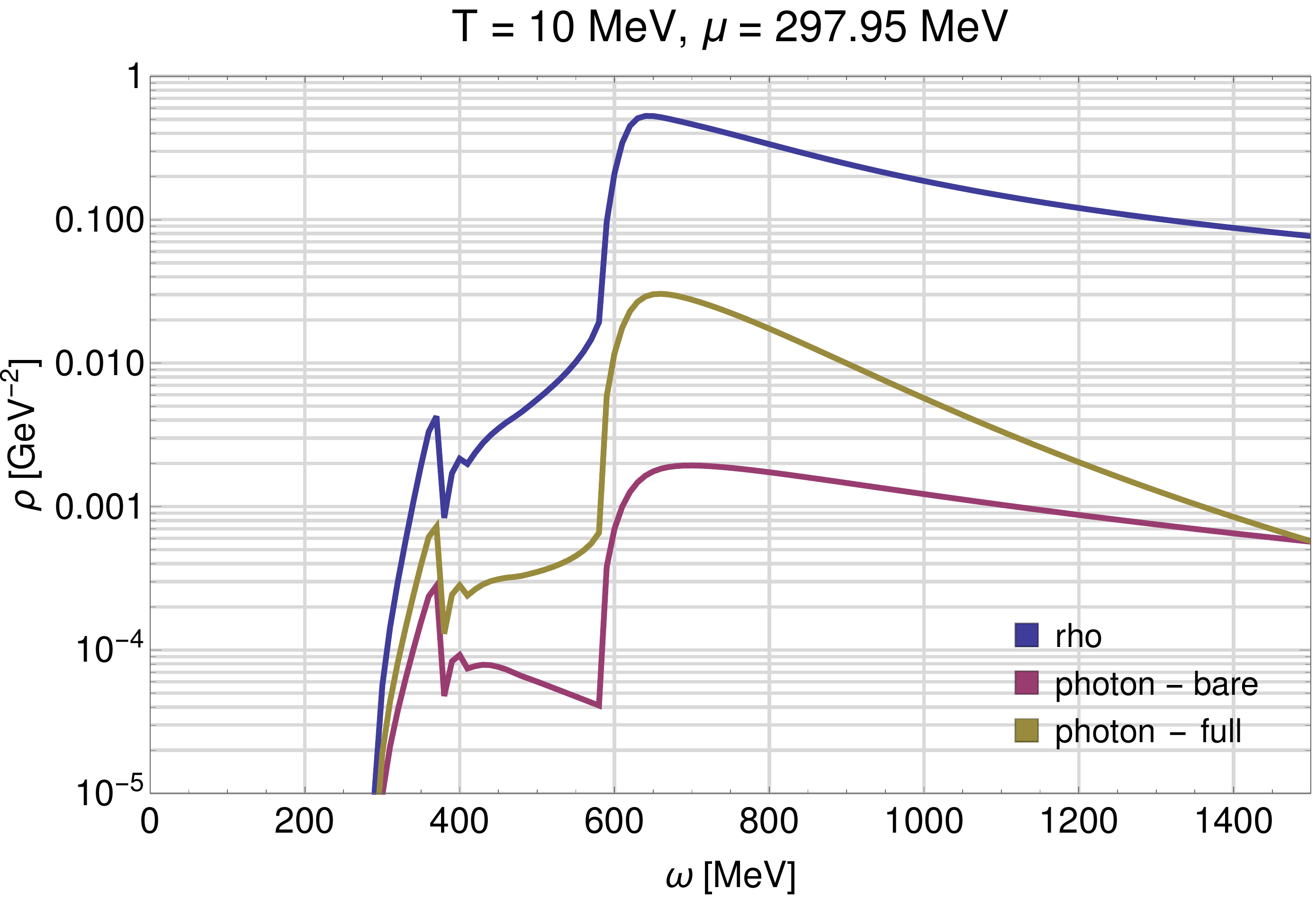}}
	\caption{The bare $\rho$, the bare photon and the full photon 2-point function in the VMD approximation, cf.~Eq.~(\ref{eq:matrix}), for different temperatures and chemical potentials, \cite{JungTanjiTripoltEtAl}. With increasing temperature, the quark-antiquark threshold moves to smaller energies while close to the CEP a non-trivial enhancement near the pion-pion threshold is observed.}
	\label{fig:EMspectral}
\end{figure}

The resulting spectral functions are shown in Fig.~\ref{fig:EMspectral} for different temperatures and (quark-)chemical potentials. In the present truncation, the $\rho$ and the photon can only decay into two pions or into a quark-antiquark pair. The thresholds for these decay channels are located at $\omega\approx 300$~MeV and at $\omega\approx 600$~MeV, respectively, for $T=10$~MeV and $\mu=0$. With increasing temperature, the pions become heavier while the quarks become lighter, which eventually gives rise to the `melting' of the spectral functions. Close to the CEP, which is located at $\mu\approx 298$~MeV and $T\approx 10$~MeV, we observe a small enhancement of the spectral functions near the pion threshold while otherwise they resemble their vacuum structure.

\begin{figure}[t!]
	{\includegraphics[width=0.49\textwidth]{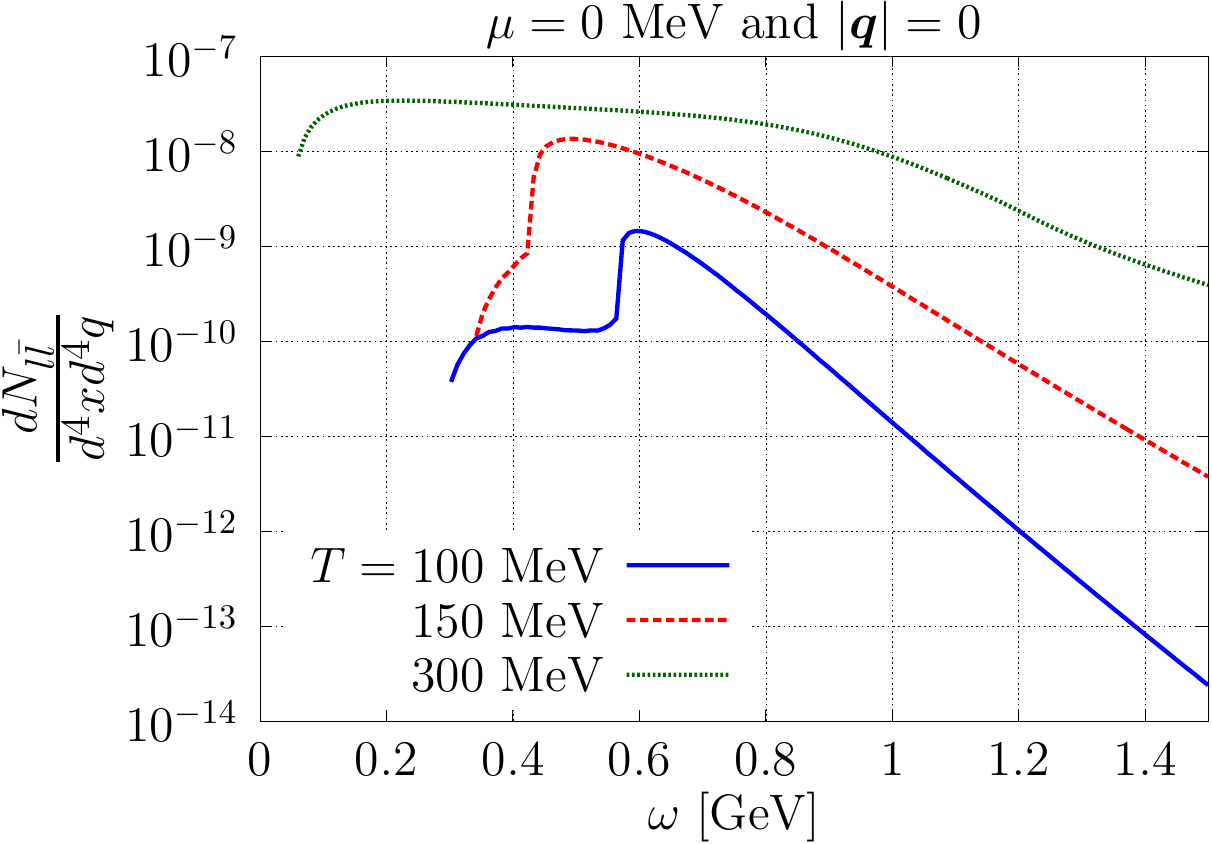}}
	{\includegraphics[width=0.49\textwidth]{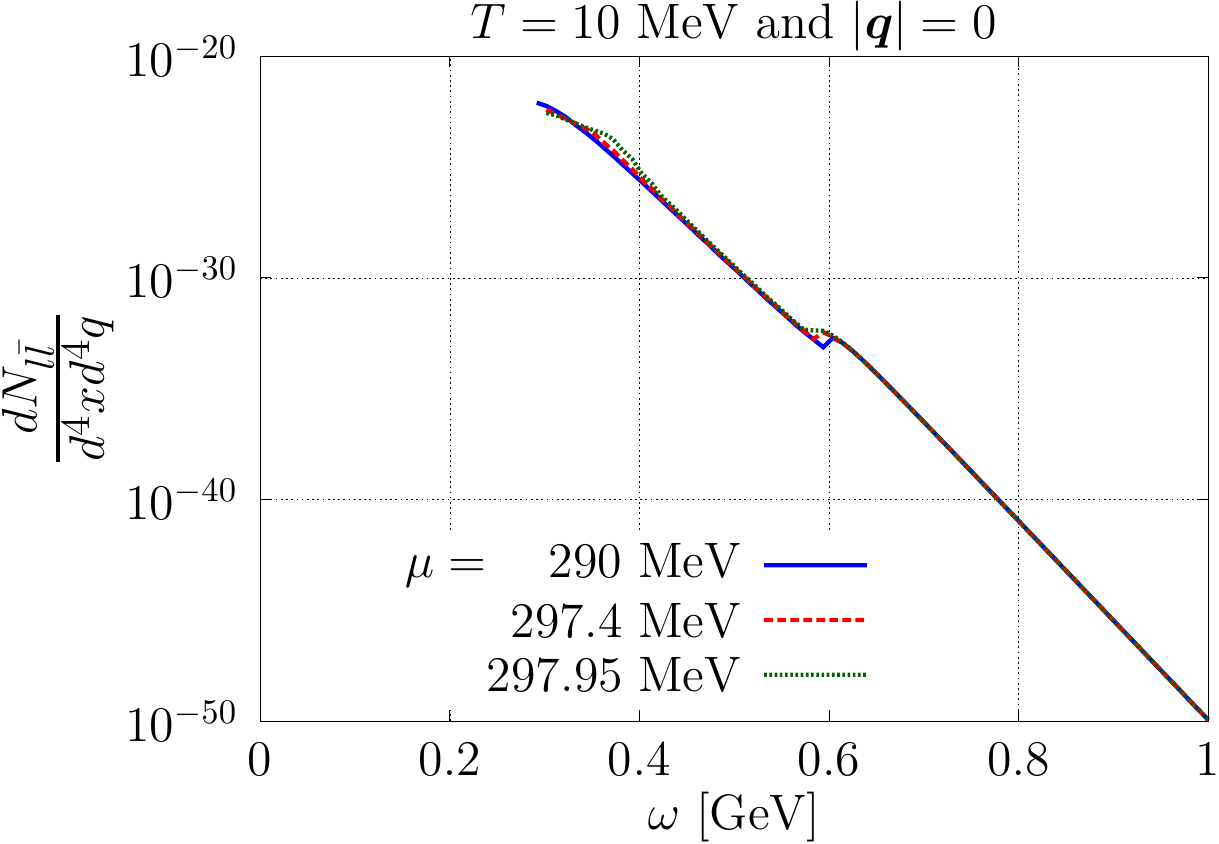}}
	\caption{Dilepton rate for different temperatures (left) and chemical potentials (right), \cite{JungTanjiTripoltEtAl}. While there is a strong dependence on the temperature, due to the in-medium modifications of the EM spectral function, cf.~Fig.~\ref{fig:EMspectral}, the effect of the CEP is almost negligible in the present truncation.}
	\label{fig:dileptons}
\end{figure}

The EM spectral function is used as input for the calculation of the dilepton rate which is obtained using the Weldon formula \cite{Weldon1990},
\begin{equation}
\frac{d^8N_{l\bar{l}}}{d^4x\,d^4q}=\frac{\alpha}{12\pi^3}
\left(1+\frac{2m^2}{q^2}\right)
\left(1-\frac{4m^2}{q^2}\right)^{1/2}
q^2 (2\rho_T+\rho_L)\, n_B(q_0),
\end{equation}
where $\alpha$ is the fine-structure constant, $m$ the lepton mass, and $n_B$ the bosonic occupation number. It expresses the dilepton production rate per space-time volume $d^4x$ and per 4-momentum interval $d^4q$ in terms of the longitudinal and transverse EM spectral function in a thermal medium. First results for the dilepton rate are shown in Fig.~\ref{fig:dileptons} for the simplifying assumptions of massless dileptons, $m=0$, and vanishing spatial momentum, i.e.~$\rho_T=\rho_L=\rho_{\tilde{A}\tilde{A}}$. 

\section{Summary}
\label{summary}
In these proceedings we present first results for the EM spectral function as well as for the dilepton rate based on an FRG framework. This framework involves an analytic continuation from Euclidean to real energies and is used to calculate the $\rho$ and the $a_1$ spectral function within an extended linear-sigma model including quarks. Non-trivial in-medium modifications are observed in the EM spectral function which serves as input for the calculation of the dilepton production rate. While a strong temperature-dependence of the dilepton rate is observed, the effects of the CEP are minimal and call for improved truncations. This will involve the inclusion of baryonic degrees of freedom and of 4-particle processes which will allow for a more realistic calculation of dilepton rates and a possible identification of signatures of the chiral crossover and the CEP.	

\section*{Acknowledgments}
This work was supported by the Deutsche Forschungsgemeinschaft (DFG) through the grant CRC-TR 211 ``Strong-interaction matter under extreme conditions''.

\bibliographystyle{elsarticle-num}

\begin{thebibliography}{}
\expandafter\ifx\csname url\endcsname\relax
  \def\url#1{\texttt{#1}}\fi
\expandafter\ifx\csname urlprefix\endcsname\relax\def\urlprefix{URL }\fi
\expandafter\ifx\csname href\endcsname\relax
  \def\href#1#2{#2} \def\path#1{#1}\fi

\end{thebibliography}


\begin{thebibliography}{10}
	\expandafter\ifx\csname url\endcsname\relax
	\def\url#1{\texttt{#1}}\fi
	\expandafter\ifx\csname urlprefix\endcsname\relax\def\urlprefix{URL }\fi
	\expandafter\ifx\csname href\endcsname\relax
	\def\href#1#2{#2} \def\path#1{#1}\fi
	
	\bibitem{RappWambachHees2010}
	R.~Rapp, J.~Wambach, H.~van Hees, {The Chiral Restoration Transition of QCD and
		Low Mass Dileptons}, Landolt-Bornstein 23 (2010) 134.
	\newblock \href {http://arxiv.org/abs/0901.3289} {\path{arXiv:0901.3289}},
	\href {http://dx.doi.org/10.1007/978-3-642-01539-7$\_$6}
	{\path{doi:10.1007/978-3-642-01539-7$\_$6}}.
	
	\bibitem{Kamikado2014}
	K.~Kamikado, N.~Strodthoff, L.~von Smekal, J.~Wambach, {Real-Time Correlation
		Functions in the O(N) Model from the Functional Renormalization Group},
	Eur.Phys.J. C74 (2014) 2806.
	\newblock \href {http://arxiv.org/abs/1302.6199} {\path{arXiv:1302.6199}},
	\href {http://dx.doi.org/10.1140/epjc/s10052-014-2806-6}
	{\path{doi:10.1140/epjc/s10052-014-2806-6}}.
	
	\bibitem{Tripolt2014}
	R.-A. Tripolt, N.~Strodthoff, L.~von Smekal, J.~Wambach, {Spectral Functions
		for the Quark-Meson Model Phase Diagram from the Functional Renormalization
		Group}, Phys.Rev. D89 (2014) 034010.
	\newblock \href {http://arxiv.org/abs/1311.0630} {\path{arXiv:1311.0630}},
	\href {http://dx.doi.org/10.1103/PhysRevD.89.034010}
	{\path{doi:10.1103/PhysRevD.89.034010}}.
	
	\bibitem{Tripolt2014a}
	R.-A. Tripolt, L.~von Smekal, J.~Wambach, {Flow equations for spectral
		functions at finite external momenta}, Phys.Rev. D90~(7) (2014) 074031.
	\newblock \href {http://arxiv.org/abs/1408.3512} {\path{arXiv:1408.3512}},
	\href {http://dx.doi.org/10.1103/PhysRevD.90.074031}
	{\path{doi:10.1103/PhysRevD.90.074031}}.
	
	\bibitem{TripoltSmekalWambach2016a}
	R.-A. Tripolt, L.~von Smekal, J.~Wambach, {Spectral functions and in-medium
		properties of hadrons}\href {http://arxiv.org/abs/1605.00771}
	{\path{arXiv:1605.00771}}.
	
	\bibitem{JungRenneckeTripoltEtAl2017}
	C.~Jung, F.~Rennecke, R.-A. Tripolt, L.~von Smekal, J.~Wambach, {In-Medium
		Spectral Functions of Vector- and Axial-Vector Mesons from the Functional
		Renormalization Group}, Phys. Rev. D95~(3) (2017) 036020.
	\newblock \href {http://arxiv.org/abs/1610.08754} {\path{arXiv:1610.08754}},
	\href {http://dx.doi.org/10.1103/PhysRevD.95.036020}
	{\path{doi:10.1103/PhysRevD.95.036020}}.
	
	\bibitem{1960AnPhy}
	J.~J. {Sakurai}, {Theory of strong interactions}, Annals of Physics 11 (1960)
	1--48.
	\newblock \href {http://dx.doi.org/10.1016/0003-4916(60)90126-3}
	{\path{doi:10.1016/0003-4916(60)90126-3}}.
	
	\bibitem{Lee:1967ug}
	B.~W. Lee, H.~T. Nieh, {Phenomenological Lagrangian for field algebra, hard
		pions, and radiative corrections}, Phys. Rev. 166 (1968) 1507--1515.
	\newblock \href {http://dx.doi.org/10.1103/PhysRev.166.1507}
	{\path{doi:10.1103/PhysRev.166.1507}}.
	
	\bibitem{JungSmekal}
	C.~Jung, L.~von Smekal, Fluctuating vector-mesons within the functional
	renormalization group, work in progress.
	
	\bibitem{JungTanjiTripoltEtAl}
	C.~Jung, N.~Tanji, R.-A. Tripolt, L.~von Smekal, J.~Wambach, work in progress.
	
	\bibitem{Weldon1990}
	H.~A. Weldon, {Reformulation of finite temperature dilepton production}, Phys.
	Rev. D42 (1990) 2384--2387.
	\newblock \href {http://dx.doi.org/10.1103/PhysRevD.42.2384}
	{\path{doi:10.1103/PhysRevD.42.2384}}.
	
\end{thebibliography}

\end{document}